\documentclass[11pt]{article}
\usepackage{graphicx}
\usepackage{placeins}
\usepackage{float}
\usepackage{subfigure}
\usepackage[toc,page]{appendix}

\usepackage{amsmath,amssymb}
\usepackage{epsfig}
\begin{document}
\title{{\bf{\Large Interior volume of ($1+D$) dimensional Schwarzschild black hole}}}
\author{ 
{\bf {\normalsize Nilanjandev Bhaumik}$
$\thanks{E-mail: nilanjandevbhaumik@gmail.com}} \,\ and \,\ 
 {\bf {\normalsize Bibhas Ranjan Majhi}$
$\thanks{E-mail: bibhas.majhi@iitg.ernet.in}}\\ 
{\normalsize Department of Physics, Indian Institute of Technology Guwahati,}
\\{\normalsize Guwahati 781039, Assam, India}
\\[0.3cm]
}

\maketitle

\begin{abstract}
  We calculate the maximum interior volume, enclosed by the event horizon, of a ($1+D$)-dimensional Schwarzschild black hole. Taking into account the mass change due to Hawking radiation, we show that the volume increases towards the end of the evaporation. This fact is not new as it has been observed earlier for four dimensional case. The interesting point we observe is that this increase rate decreases towards the higher value of space dimensions $D$; i.e. it is a decelerated expansion of volume with the increase of spacial dimensions. This implies that for a sufficiently large $D$, the maximum interior volume does not change. The possible implications of these results are also discussed.    
\end{abstract}

\section{Introduction}
   Unlike the horizon surface area of a black hole, until some very recent developments, interior volume was not a subject of popular interest in black hole thermodynamics. The concept of interior volume in case of black holes contains some implicit difficulty in understanding. Since in curved geometry simultaneity surface looses its significance to be taken as the volume, we need to try to find an alternative definition of volume. Eventually multiple ways to define volume in curved geometry have been proposed \cite{Parikh:2005qs}--\cite{Gibbons:2012ac}. Recently, Christodoulou and Rovelli suggested the volume to be the largest space like surface with spherical nature inside the black hole horizon \cite{Christodoulou:2014yia}. From now we call this as CR volume. They showed that the maximum interior volume for a Schwarzschild case, is linearly proportional to advanced time. The same has been concluded for Kerr metric also \cite{Bengtsson:2015zda}. Their formulation was completely classical. More recent analysis showed that this method carries vast significance while considering from thermodynamical point of view. It turned out that the CR formulation is useful to estimate phase space volume as well as the horizon entropy of a black hole \cite{Zhang:2015gda}. Moreover, such a construction of volume, when Hawking radiation \cite{Hawking:1974sw} is taken into account, has significant importance to resolve the information paradox problem (see \cite{Ong:2015tua}--\cite{Ong:2016xcq}, for details). It has been observed that the volume is always increasing with advanced time and therefore we shall have a large amount of volume at the end of the evaporation \cite{Ong:2015dja}. Hence there is enough place to store the information. This has been further elaborated and clarified in \cite{Christodoulou:2016tuu}. Motivated by these facts and implications, here, in this paper we shall mainly concentrate on the definition of CR volume for higher dimensional Schwarzschild black hole.

   Our aim, in this paper, is to see how the CR volume behaves for higher dimensional black holes. Due to simplicity and availability of the relevant quantities, we shall restrict our analysis on a ($1+D$)-dimensional Schwarzschild black hole.
   Taking into account the mass change due to Hawking radiation, it will be observed that the rate of change of maximum volume with advanced time for a particular value of spacial dimension is positive. Therefore, like the ($1+3$)-dimensional case \cite{Ong:2015dja}, it is increasing. But the important observation, we shall note that, this rate is decreasing with the increase of dimensions of space. At sufficiently large value of $D$, the change is negligible and hence the maximum volume remains same. This will be shown by plotting the rate of change with $D$. Such fact will be further bolstered by calculating the total change of CR volume for the total evaporation time. We shall see that this is going to vanish in limit of $D\rightarrow\infty$.

   The organization of the paper is as follows. In section \ref{D}, we present out main analysis and results. Finally, we discuss our results and implications in section \ref{discuss}.

\section{\label{D}($1+D$) dimensional Schwarzschild black hole}
   Here the dependence of the CR volume on the spacetime dimensions will be extensively discussed for an arbitrary dimensional Schwarzschild black hole. Considering the Hawking evaporation, we shall observe that at large dimensions the volume almost remains constant with respect to the advanced time.

         The  ($1+D$) dimensional Schwarzschild black hole metric is given by 
\begin{equation}
ds^2 = -f(r)dt^2 + \frac{dr^2}{f(r)}+r^2d\Omega^2_{(D-1)}~,
\label{metric}
\end{equation}
where $f(r)=1-(r_H/r)^{D-2}$ with the horizon radius is $r_H=\Big[\frac{16\pi M}{(D-1)A_{D-1}}\Big]^{\frac{1}{D-2}}$. Here $M$ is the mass of the black hole and $A_{D-1}=\frac{2\pi^{D/2}}{\Gamma(D/2)}$. The horizon area is found to be 
\begin{equation}
\mathcal{A} =  A_{D-1} {r_H}^{D-1} ~.
\label{area}
\end{equation}

      To use the CR formulation \cite{Christodoulou:2014yia}, first of all we need to transform the metric (\ref{metric}) in Eddington-Finkelstein co-ordinates $( v,r,\theta, \phi)$: 
\begin{equation}
ds^2=-f(r)dv^2+2dvdr+r^2d\Omega^2_{(D-1)}~,
\label{1.02}
\end{equation}
where $v=t+r^*$ with $dr^*=dr/f(r)$.
Now the maximum volume inside the horizon is defined as the value of the maximum of the volume of a $D$-dimensional surface $\Sigma$, which is a direct product of a ($D-1$)-sphere and a curve on the ($v-r$)-plane; i.e. 
$\Sigma=\gamma \times S^{(D-1)}$.
According to \cite{Christodoulou:2014yia}, the maximum volume corresponds to $r=$constant surface. Then from (\ref{1.02}), the proper volume of this surface is given by
\begin{eqnarray}
V&=&\int_V \sqrt{-g} ~\,d\Omega'dv 
\nonumber
\\
&=&\int_V \sqrt{r^{2(D-1)}f(r)} ~\,d\Omega'dv
\label{new1}
\end{eqnarray}
Now inside the horizon the metric coefficient is given by $f(r)=(r_H/r)^{D-2}-1$ and so the integration over the angular coordinates can be done. Therefore the volume turns out to be
\begin{equation}
V = A_{D-1}\int^{v}r^{D-1}\sqrt{\Big(\frac{r_H}{r}\Big)^{D-2}-1}~dv~.      
\label{1.0.5n}         
\end{equation}
Then the maximum volume inside the horizon will be given by the value of above integration for a particular $r$ at which the integrand is maximum. It is found that this value of $r$ is $r_c=r_H(\frac{D}{2(D-1)})^{\frac{1}{D-2}}$. One can check that for $D=3$ its value matches with the earlier analysis \cite{Christodoulou:2014yia}, done for usual ($1+3$) dimensional case. Substituting this in (\ref{1.0.5}) we obtain the maximum volume inside the horizon as   
\begin{equation}
V = A_{D-1}\Big(\frac{D}{2(D-1)}\Big)^{\frac{D-1}{D-2}}\Big(\frac{D-2}{D}\Big)^{1/2}\int^{v}r_H^{D-1}dv~.      
\label{1.0.5} 
\end{equation}
Note that in the above the $r_H$ term can not be kept outside the integration as $r_H$ can change with the advanced time $v$ if we consider the Hawking radiation. In the below we shall consider the Hawking evaporation of the black hole, in which case the mass is changing with $v$.

\begin{figure}[H]
  \centering
    \includegraphics[width=0.4\textwidth]{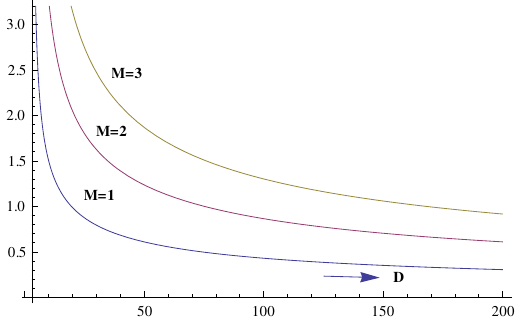}
     \caption{{\it{$\frac{dV}{dv}$ vs $D$ plot.}}}
     \label{Figure2}
\end{figure}

   Substituting the value of $r_H$ in terms of mass $M$, one obtains the variation of the CR volume with respect to the Eddington-Frinkelstein time as
\begin{eqnarray}
\frac{dV}{dv}=L(D)M^{\frac{D-1}{D-2}}~,
\label{volume1}
\end{eqnarray}
where $L(D)=(\frac{D}{2(D-1)})^{\frac{D-1}{D-2}}(\frac{D-2}{D})^{1/2}(\frac{16\pi}{D-1})^{\frac{D-1}{D-2}}A_{D-1}^{\frac{1}{2-D}}$. Note that for $D\geq 3$ the right hand side of the above is always positive. This implies that the interior volume is always increasing, even if the mass of the black hole is decreasing. The same was also concluded in \cite{Ong:2015dja} for ($1+3$) dimensional Schwarzschild black hole. {\it Therefore we can say that this feature is independent of the dimensions of spacetime}. The new observation, for the present case, is the following. Here the value of right hand side depends not only the value of mass, but also on the space dimensions $D$. Now if one plots this as a function of $D$ with a fixed value of mass, then the nature of $dV/dv$ is the following (See Figure \ref{Figure2}). 
\noindent
This shows that the rate of increase of volume decreases with the increase of spacetime dimensions; i.e. the volume increase decelerates.  It implies that at sufficiently large $D$, the increment of interior volume is almost negligible.  This fact will also be again discussed in the next part of paper by calculating the total volume change for the time interval of total evaporation by Hawking process.

    Now if we consider only Hawking radiation as the cause of change in black hole mass, then the flux is given by \cite{Hod:2011zzb}
\begin{eqnarray}
\frac{dM}{dv}=-\frac{B(D)}{r_H^2}~,
\label{flux} 
\end{eqnarray}
where $B(D)=\frac{(D+1)(D-2)}{2} (\frac{D-2}{4\pi})^{D+1} (\frac{D}{2})^{\frac{D-1}{D-2}} (\frac{D}{D-2})^{\frac{D-1}{2}} \frac{D\zeta(D+1)\hbar}{\pi}$. The expression has been obtained by Eikonal approximation and is valid for large frequency; i.e. in the limit $D>>1$.  
If the initial black hole mass is $M_0$, then after time $v$ the expression for black hole mass can be obtained by integrating (\ref{flux}). It turns out to be
\begin{eqnarray}
M(v)=\Big(-P(D)v+{M_0}^{\frac{D}{D-2}}\Big)^{\frac{D-2}{D}}~,
\label{mass}
\end{eqnarray} 
with $P(D)=B(D) (\frac{(D-1)A_{D-1}}{16\pi})^{\frac{2}{D-2}} \frac{D}{D-2}$.
Next, replacing the above value of black hole mass in (\ref{volume1}), we obtain the rate of volume change as a function of advance time $v$: \begin{eqnarray}
\frac{dV}{dv}=L(D)\Big(-P(D)v+{M_0}^{\frac{D}{D-2}}\Big)^{\frac{D-1}{D}}
\label{volume2}
\end{eqnarray}
Also note that if the complete evaporation time is $v_{max}$ by Hawking process, then its value is can be found out by setting $M=0$ in (\ref{mass}). It is given by
 \begin{eqnarray}
v_{max}=\frac{M_0^{\frac{D}{D-2}}}{P(D)}~.
\label{vmax}
\end{eqnarray}
Within this time, the total volume change, calculate from (\ref{volume2}), is
\begin{eqnarray}
&&\int_{V_i}^{V_f}dV=L(D)\int_{0}^{v_{max}}(-P(D)v+{M_0}^{\frac{D}{D-2}})^{\frac{D-1}{D}}dv
\nonumber
\\
&&\Rightarrow \Delta V=V_f-V_i=Q(D){M_0^{\frac{2D-1}{D-2}}}~,
\label{volume3}
\end{eqnarray}
where the value of $Q(D)$ is given by
\begin{eqnarray}
Q(D)&=&\frac{L(D)}{P(D)}\frac{D}{2D-1}
\nonumber
\\
&=&(16\pi)^{\frac{D+1}{D-2}}(\frac{1}{D-1})^{\frac{2D}{D-2}}(\frac{D-2}{D})^{D/2}(\frac{4\pi}{D-2})^{D+1}
\nonumber
\\
&\times&\frac{2\pi}{D(D+1)(2D-1)\zeta(D+1)\hbar}(A_{D-1})^{\frac{3}{2-D}}~.
\label{Q}
\end{eqnarray}
Let us now plot $\Delta V$ as a function of $D$ to reveal the nature of volume at the different dimensions.
\begin{figure}[H]
\begin{center}
\includegraphics[width=0.4\textwidth]{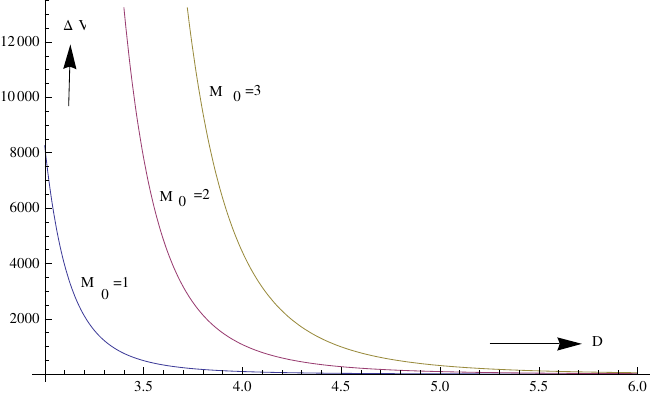}
\caption{$\Delta V$ vs $D$ plot. Here $\Delta V$ is taken in units of $\hbar$.}
\label{Figure3}
\end{center}
\end{figure}
\noindent
Figure \ref{Figure3} shows that the total volume change decreases with the increase of space dimensions. For very large value of $D$, the quantity is almost negligible. This implies that for a sufficiently higher dimensional Schwarzschild black hole the maximum interior volume remains almost same even if it is radiating through Hawking process.

\begin{figure}[H]
\begin{center}
\includegraphics[width=0.4\textwidth]{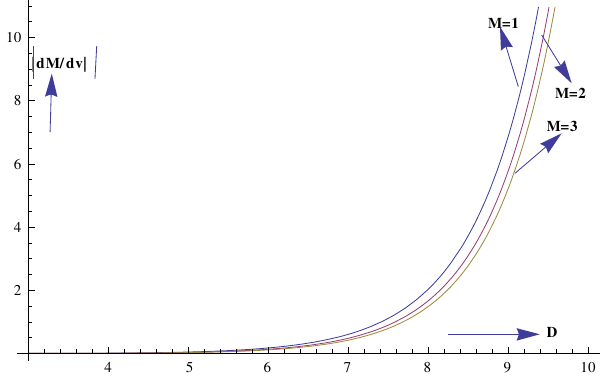}
\caption{$|dM/dv|$ vs $D$ plot. Here $|dM/dv|$ is taken in units of $\hbar$.}
\label{Figure4}
\end{center}
\end{figure}

   The possible reason behind the decelerated increase of volume can be explained as follows. If we look at the ($1+3$) dimensional case with Hawking radiation taking into account, the flux (rate of energy emission from the black hole) is increasing with the decrease of mass ($dM/dv\sim 1/M^2$) and correspondingly the rate of volume increment decreases as $dV/dv\sim M^2$; i.e. at late time the flux increases whereas the volume increases with decreased rate. The similar is also happening with the analysis in the higher dimensions. At higher dimensions the flux increases (see Figure \ref{Figure4} corresponding to Eq. (\ref{flux})) 
\noindent
and so correspondingly we obtained decelerated increase of interior volume. It implies that the increase of Hawking flux reflects a negative impact on the maximum interior volume.

\section{\label{discuss}Discussions} 
   Considering only Hawking radiation as the reason of change in black hole mass, we observed that for a ($1+D$) dimensional Schwarzschild black hole the rate of increase in CR volume is always positive, suggesting a nonzero interior volume even at the end of Hawking radiation process. This is similar to the result found for ($1+3$) dimensional case \cite{Christodoulou:2014yia,Ong:2015dja}. It is interesting to note that at the end of Hawking radiation process when mass of the black hole touches the zero value, the CR volume remains non zero, leaving a possibility of solving information loss paradox. 
    In this paper, we got an additional feature in the change of black hole interior volume. It has been observed that as we go to higher dimensions keeping the black hole mass constant, the volume increase gets decelerated. Here we have also estimated the total change in CR volume ($V_f-V_i$) during the Hawking radiation process in terms of its initial mass and we have again seen that this estimated volume change also decreases for the increase in dimensions. So from these results we can assert that if we go to very high dimensions we are supposed to get an almost negligible volume increase rate or constant CR volume of a Schwarzschild black hole. The precise physical significance of which is unknown to us.

    Now we want to bring some important facts for higher dimensional case which may have connection with our analysis. 
    From recent analysis \cite{Hod:2016rmg} it has been observed that the time difference between the emission of two consecutive quanta in Hawking radiation process decreases with the increase of number of dimensions. The typical value of $D$ for continuous spectrum was found out to be $D\geq 10$. Then one may be curious to know if there is any physical connection between the continuous emission with the decelerated rate of volume increase with increasing dimensions.
    Another observation can also be important in this direction. The perturbative analysis for non-uniform black strings suggests that for $ D>13 $ they are stable and  so these can be the end point of Gregory-Laflamme instability (see the review \cite{Reall:2015esa}). Therefore it might be possible that this nonuniform nature can be expected to offer a resistance in the rate of volume increase of the black hole. All these aspects are now under investigation.

\vskip 9mm
\section*{Acknowledgments}
The research of one of the authors (BRM) is supported by a START-UP RESEARCH GRANT (No. SG/PHY/P/BRM/01) from Indian Institute of Technology Guwahati, India.

\end{document}